\documentclass[twocolumn,floatfix,showpacs,preprintnumbers,amsmath]{revtex4}

\usepackage{psfrag}
\usepackage{graphicx}
\usepackage{amssymb,graphics}
\usepackage{color}
\usepackage{hyperref}
\usepackage{xcolor}

  \definecolor{dark-grey}{rgb}{0.2,0.2,0.2}
  \definecolor{dark-green}{rgb}{0,0.5,0}
  \definecolor{dark-orange}{rgb}{0.9,0.5,0}
  \hypersetup{colorlinks=true, linkcolor=dark-grey, urlcolor=dark-grey, citecolor=dark-grey}
  \hypersetup{colorlinks=true, urlcolor=blue}

\begin{document}

\title{Some Fundamentals of Adhesion in Synthetic Adhesives}
\author{Cyprien Gay}
\affiliation{%
Centre de Recherche Paul Pascal - CNRS, 115 av Dr Schweitzer, 33600 Pessac, France
}
\date{2003}


\pacs{
     } 
\maketitle

\newcommand{\hs}{\hspace{0.7cm}}
\newcommand{\be}{\begin{equation}}
\newcommand{\ee}{\end{equation}}
\newcommand{\bee}{\begin{eqnarray}}
\newcommand{\eee}{\end{eqnarray}}
\newcommand{\fin}{\nonumber\\}

Contact:\\ cgay@crpp.u-bordeaux.fr (2003),\\
\href{mailto:cyprien.gay@univ-paris-diderot.fr}{cyprien.gay@univ-paris-diderot.fr} (2016).\\

This article is available
in \href{http://dx.doi.org/10.1080/0892701031000068690}{Biofouling, 2003 Vol 19 (Supplement), pp 53--57}.
The present postprint version is identical to the published version.\\

{\em (Received 27 August 2002; in final form 14 November 2002)}

\section*{Abstract}

Various adhesion mechanisms that have been understood
in the field of synthetic adhesives are described and these
are linked with situations relevant to fouling issues. The
review mainly deals with mechanical aspects of adhesion
phenomena, with an emphasis on the role of the elasticity
of the bodies, called substrata, attached by adhesive. The
consequences of thin film geometry of the adhesive
material are described, such as various heterogeneous
deformations upon traction. The importance of the
bonding process is discussed, as well as some examples
of non-wetting surfaces. Some basic ideas of fracture
mechanics are provided and in particular, the behavior of
layered systems is discussed. Rolling sticky objects and
peeled (flexible) adhesive tapes display similar mechan-
isms and it is shown how they differ from the normal
separation of rigid bodies. Some issues directly related to
fouling issues are also discussed, such as forces and
torques acting on shells, the advantages of gregarious
settlement behavior and concepts for fouling release and
antifouling.

{\bf Keywords:} adhesion mechanisms; synthetic adhesives; the
bonding process; non-wetting surfaces; fracture mechanics;
fouling issues.

\begin{figure}
\includegraphics[width=\columnwidth,clip]{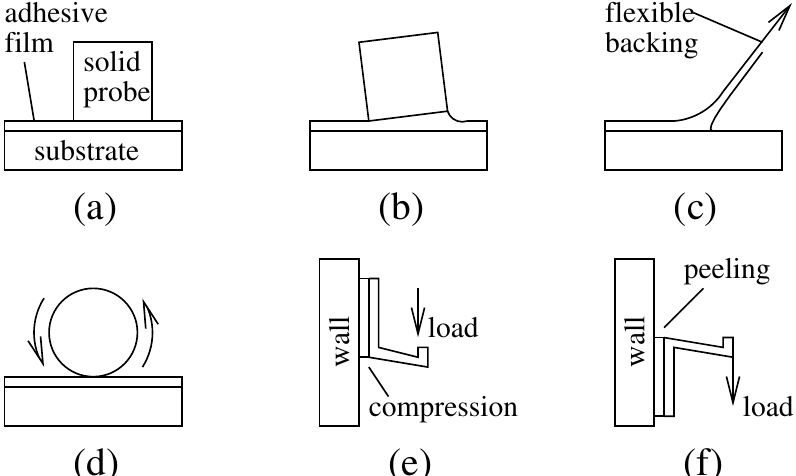}
  \caption{Influence of substratum flexibility on adhesion.
Normal separation of rigid bodies (a) is slightly stronger than
tilting (b) and much stronger than peeling (c). Rolling (d) is similar
to peeling in the separation region. The usual design of adhesive
hooks (e) avoids the tendency to peeling that would be induced by
the finite flexibility of the material in other geometries (f).}
  \label{fig1}
\end{figure}

\section*{THIN FILMS}

Synthetic adhesives are usually soft, essentially
incompressible materials that do not flow. They are
used in the form of thin films of typically 100 m m in
thickness. Their adhesive properties are often tested
in a controlled test that mimics the typical situation
of two solid bodies linked by an adhesive. In the
so-called probe-tack geometry, introduced by Zosel
(1985), a flat, solid punch, called the probe, is brought
into contact with an adhesive film deposited on a
rigid substratum. After a certain contact time, the
force is recorded while the probe is being pulled
away (Figure 1a). Because the adhesive material is
almost incompressible, pulling the solid bodies apart
causes a strong convergent deformation and thus a
strong resistance to separation, i.e. to good adhesion.
Furthermore, because the resistance of the adhesive
is so important, the deformability of the entire
system, and in particular the compliance of the
machine itself may play a role. Elastic energy is
stored for some time, until the adhesive gives way
and causes a sudden separation, as discussed in a
specific geometry by Francis et al. (2001). Sudden
separation is possible when instabilities develop in
the bulk of the adhesive material so as to relieve
much of the stress. Such instabilities have been
observed directly during separation in a modified
version of the probe-tack test (Lakrout et al., 1999)
and shown to fall into two main categories, viz.
fingering instabilities and cavitation. Both types of
instabilities are driven by the need for relieving the
stress.

When air fingers protrude from the edge of the
sample towards the center, they bring atmospheric
pressure well into the sample and thus relieve the
negative pressure that has developed near the center
due to the applied traction. In the case of purely
viscous liquids, this is the well-known Saffman-
Taylor instability (Saffman et al., 1958). Cavitation
also relieves stress since the growing bubbles
provide some of the extra volume required by the
plate separation. Since adhesive materials are
usually more elastic than viscous, they are most
often subject to cavitation, which involves strong
deformations on the scale of the sample thickness.
They almost never undergo fingering, which would
involve deformations on the scale of the entire
sample. When a very viscous liquid is used in this
geometry instead of proper adhesive material,
fingering is observed at low separation rates while
cavitation occurs at higher rates (Poivet et al.,
unpublished observations), which further illustrates
the fact that both mechanisms compete in relieving
the applied tensile stress. Both fingering and
cavitation involve strong deformations in the sample
during separation and thus cause a large energy
dissipation, ranging typically from $100$ to $1000$ J/m$^{-2}$
for good adhesives, which is much higher than
typical surface energies in the $0.01--0.1$ J/m$^{-2}$ range.

A number of properties are associated with an
adhesive. It is usually a soft material that does not
flow and made of polymers whose molecular
architecture may vary (cross-linked polymers,
block-copolymers). It can accommodate large defor-
mations in order to dissipate a large amount of
energy, yet it is essentially incompressible.

The need for relieving the applied stress is so
strong in the thin film geometry that if the adhesive
material is too resistant, e.g. if it is very elastic,
fingering and cavitation from the edge of the sample
may occur at the interface, as shown by Ghatak
(2000) and Shull (2000). This highlights the fact that
not only the bulk of the adhesive film, but also its
interfaces with the solid bodies set up during the
bonding process, must resist separation efficiently.

\begin{figure}
\includegraphics[width=\columnwidth,clip]{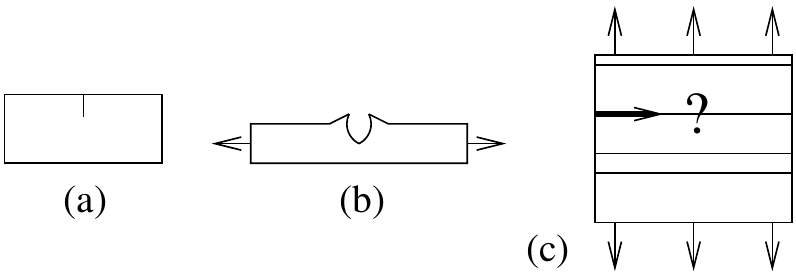}
  \caption{Adhesion phenomena and fracture mechanics. Elastic
band with a notch at rest (a) and under tension (b). In a layered
system under traction (c), contrasts in the elastic properties and
thicknesses of the different layers may cause the fracture to
migrate from its initial location to another one, either within a
material (cohesive rupture) or at the interface between them
(adhesive rupture).}
  \label{fig2}
\end{figure}

\section*{THE BONDING PROCESS}

The quality of the contact between the adhesive and
the substrata results from the interplay of inter-
actions at different length scales. On the molecular
scale, it results generically from van der Waals
interactions (Israelachvili, 1992). Surface chemical
bonds (Gent et al., 1972), macromolecular interdigita-
tion (Rapha\"el et al., 1992) or macromolecule
elongation (Lake et al., 1967), however, may enhance
significantly the strength of the interface for specific
substratum-adhesive pairs. Surface treatments and
cleaning are also essential. On the micrometre scale,
solid substrates usually display some degree of
surface roughness (Greenwood et al., 1966) that may
reduce the degree of intimacy of the contact with the
adhesive if the adhesive material is not very soft
(Dahlquist, 1969; Fuller et al., 1975; Creton et al., 1996;
Crevoisier et al., 1999). For very soft adhesives,
however, the surface roughness of the substratum
may paradoxically enhance the strength of the
interface, as small air bubbles trapped at the interface
may generate suction effects upon traction (Gay \&
Leibler, 1999b).

When seeking a bad adhesion even with good
adhesives, the main strategy is to weaken the contact
at the body surface, and in particular to use
non-wetting surfaces, i.e. surfaces such that the
gain in surface energy upon making contact with
the adhesive is small. When spread on such surfaces,
most liquids do not spread spontaneously but gather
into droplets. On very weakly wetting surfaces, in
particular on lotus leaves and on other plants and on
some insect wings (Wagner et al., 1996; Barthlott et al.,
1997; Neinhuis et al., 1997), millimetric drops are
almost complete spheres and can easily roll. Solid
surfaces were successfully designed to mimic this
“lotus effect” and the surface design, an array
of micrometer-sized poles, enhances the weak
wettability achieved through chemical treatment
(Bico et al., 1999).

\begin{figure}
\includegraphics[width=\columnwidth,clip]{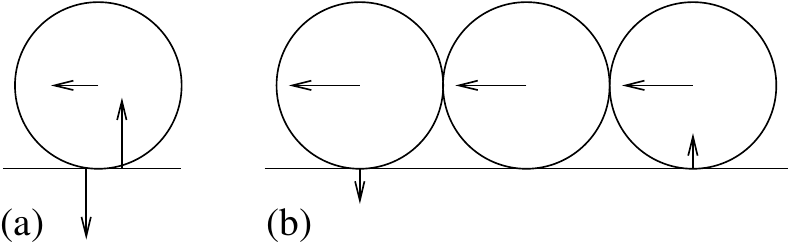}
  \caption{Advantage of gregarious settlement. Organisms on a
ship’s hull are subjected to a hydrodynamic force (drawn
horizontally for simplicity). They transmit not only this force to
the hull (not shown), but also the resulting torque (drawn as two
vertical arrows). In the case of a single organism (a), the torque
involves large compressive and tensile stresses when the contact
region with the hull is narrow. If two or more organisms are rigidly
bound to one another (b), the torque that they transmit to the hull
involves much smaller stresses in the contact regions, thus
impeding detachment.}
  \label{fig3}
\end{figure}

\section*{FRACTURE MECHANICS}

The mechanisms described above, by which the
adhesive material undergoes heterogeneous defor-
mations upon traction, involve the bulk of the
adhesive film. However, as seen from a distance, the
adhesive joint appears to be the interface between
both solid bodies, and its deformations due to
traction can be seen as a fracture propagating at this
interface. In this perspective, adhesion phenomena
are connected to the field of fracture mechanics.
A small notch cut on one side of an elastic band
(Figure 2a) illustrates a number of points. If the band
is pulled gently, it stretches homogeneously except in
the vicinity of the notch where it is less stretched
(Figure 2b). In other words, elastic energy is stored in
the material but the notch relieves some of this
energy. If the band was cut further, the notch would
be longer and relieve more stress. This will not
happen spontaneously unless it is pulled more
strongly. It will generally happen when the elastic
energy released by the notch propagation is
sufficient to break further bonds at the interface
and propagate the fracture. This is a central concept
in fracture mechanics, called Griffith’s criterion
(Griffith, 1920).

Since the energy stored depends on the elastic
properties of the entire system, not only does the
elasticity of the testing apparatus play a role if it is
soft enough, but very non-intuitive behaviors may
arise. The fracture may propagate within the
adhesive film i.e. its cohesion affected, or it may
take place at the interface with one of the
substrates. More generally, when considering a
stack of several layers from different materials
(Figure 2c), the system may break upon traction or
peeling. The system may choose between different
fracture mechanisms (interfacial or cohesive, with
further choice in a layered system). At first, it
usually chooses the weakest mechanism in terms of
force since it triggers separation first and thereby
relieves the stress on the other possible mechan-
isms. In the long run, however, fracture propagation
is driven by the elastic energy stored in the system
under tension which must exceed the energy
required by the fracture. The system thus tends to
choose the separation mechanism that dissipates
the smallest amount of energy per unit surface area
since the fracture will then be able to propagate
faster.

But, when a sheet of paper is torn out of note-pad,
depending on how exactly the paper is pulled and
how the note-pad is held, the sheet may detach
neatly or be torn. Due to the mismatch in the elastic
properties of the note-pad and the paper sheet, the
fracture may propagate along a path that does not
minimize the energy dissipation. More generally, in a
layered system, the elastic properties of the various
layers may prevent the fracture from migrating
towards its energetically optimal location.

\begin{figure}
\includegraphics[width=\columnwidth,clip]{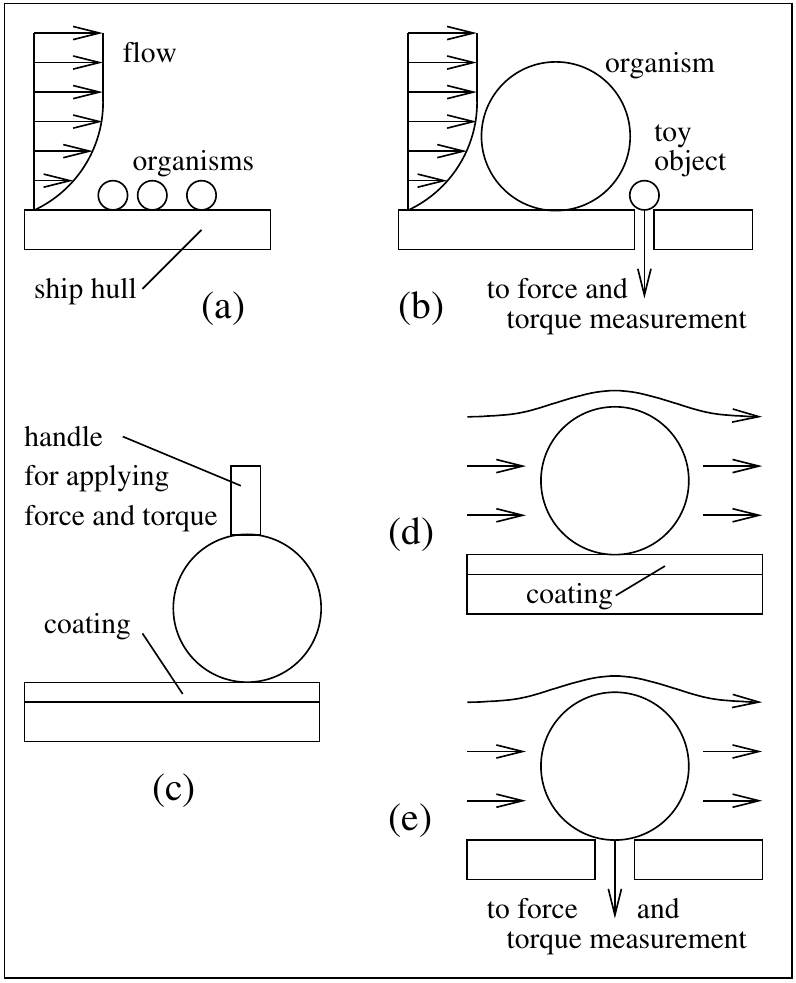}
  \caption{Schematic comparison between ship-scale (a) and
laboratory-scale experiments (b-e). The size of real organisms is
some small fraction of the boundary layer under real ship
conditions (a). In order for laboratory experiments to determine
the hydrodynamic forces involved (which include both drag and
lift) and the resulting torque transmitted to the hull, the organisms
themselves should not be used. Instead, small ‘toy’ objects should
be used so that their size is the same small fraction of the
laboratory-scale boundary layer (b). After proper upscaling of the
measured stresses to the size of the organisms, fouling-release
coatings can then be tested by either applying the force and torque
through direct mechanical action (c) or by applying a laboratory-
scale flow (d) which has been calibrated (e).}
  \label{fig4}
\end{figure}

\section*{PEELING, TILTING AND ROLLING}

As discussed above, separating both substrata in the
probe-tack geometry is difficult because the lateral
extension of the adhesive joint is much greater than
its thickness (thin geometry). It is also essential that
both substrata are rigid (Figure 1a). Indeed, when an
adhesive tape is considered, the sticky side is the
adhesive film, while the non-sticky side is the (very
flexible) backing. Thus, the backing plays the role of
one of the substrata. But peeling a tape off a table is
quite easy because the backing is flexible and only a
restricted region of the adhesive is under tension at
any given time during peeling (Figure 1c). Peeling is,
in general, so much easier that it is avoided as much
as possible in practical situations. For instance,
adhesive hooks for bathroom or kitchen utensiles are
designed in such a way that the hook deformations
induced by a hanging weight cause the adhesive
layer to be locally compressed rather than peeled,
otherwise the adhesive joint would soon be ruined
(Figure 1e, f).

Although similar to peeling at first sight, detach-
ment by tilting is difficult if both the substrata are
rigid. Indeed, tilting one object (for instance, the
probe in the probe-tack geometry) involves com-
pression of the adhesive film along the vicinity of one
edge, and traction on the other edge and on most of
the region of contact with the object (Figure 1b).
Hence, tilting the object is very similar to pulling it.

Tilting becomes similar to peeling only at the point
where the adhesive joint is not longer thin, i.e. if its
lateral dimensions are comparable to its thickness, it
is then easy. This feature is used by tropical lizards
(geckoes), which can climb walls at considerable
velocities. Their adhesive feet consist of a hierarchy
of structures. The smallest structure is the spatula
(0.2 $\mu$m in width). Adherence to the wall is
important, even if it is rough, because spatulae can
achieve contact with it somewhat independently of
one another. However, foot removal is obviously
achievable without too much effort. Because spatulae
are independent, they can be tilted simultaneously
and peeled away from the wall. Due to their small
lateral dimension, this can be done with little effort
(Autumn et al., 2000). In short, geckos use peeling on
the lowest scale to lift their feet.

A round object can roll on a sticky surface,
whereby a new contact with the adhesive is made at
the front, while separation, similar to peeling, is
achieved at the rear (Figure 1d). Because the
separation dissipates more energy than is provided
by the contact, a sufficient sideward force must be
applied for the object to move. Experiments
conducted on glass cylinders in contact with rubber
(Charmet et al., 1995 and references therein) show
that such rolling can occur also if the applied force
has a normal component. For example, cylinders can
roll both on and under an inclined rubber sheet.
Similarly, marine organisms attached to a boat’s hull
undergo a sideways force as well as a lift force
(because the flow is usually in the inertial rather than
viscous regime). They may thus roll if the water
velocity is moderate, and they may even detach if the
lift component of the force is high. If the organisms
are rigid, gregarious settlement behavior hinders
flow-driven detachment, since when attached
together, they constitute a solid with a large lateral
extension, which turns easy rolling into more
demanding tilting (Figure 3).

\begin{figure}
\includegraphics[width=0.4\columnwidth,clip]{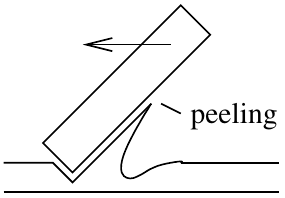}
  \caption{Schematic separation mode of an extended rigid object
(such as aggregated rigid organisms) from a coating consisting of a
flexible, though weakly extensible, thin upper layer and a much
softer lower layer. Deformations are magnified for clarity.}
  \label{fig5}
\end{figure}

\section*{FOULING ISSUES}

Testing fouling release properties of coatings some-
times involves real-size experiments on ships, but
more usually laboratory scale characterization.
Hydrodynamic effects scale with dimension. In
particular, boundary layers are much smaller in
laboratory tests than under ship-operating con-
ditions (Schultz et al., 2003), thus attached bodies
should be scaled down accordingly (Figure 4a, b).
The interpretation of laboratory tests therefore needs
to combine information from different types of
experiments. For example, small ‘toy’ particles on
small ships can be used in laboratory tests to mimic
organisms to measure what lateral and normal drag
forces and torques they undergo at various locations
on the ship and at various equivalent ship speeds
(Figure 4b). After proper upscaling of these forces
and torques, the laboratory or small boat tests with
live organisms should be performed under con-
ditions where forces and torques can be reproduced
(Figure 4d). These parameters can be measured if the
organisms are located on small, instrumented
patches inside larger panels (Figure 4e). Alterna-
tively, if the organism under study is rigid, the forces
and torques can be applied to it mechanically, in the
absence of any flow (Figure 4c). If the organism is
much more rigid than the coating, or, more generally,
if its mechanical properties are known, its detach-
ment can be mimicked with more convenient real-
size ‘toy’ objects with an experimental set-up that
controls the forces and torques.

The mechanics of adhesion briefly described in the
previous sections suggest the following direction for
fouling release coatings. If an aggregate of rigid
organisms on the surface is considered on a classical
thin coating, the only detachment mode is tilting
(Figure 3). But, if the coating is flexible in some sense,
then an easier detachment mode, similar to peeling,
may appear. It is suggested that a coating consisting
in an inextensible, flexible, very thin upper layer and
a much softer lower layer, may provide enough local
compliance to accompany a tilting object and peeling
away from it, thus easing its detachment (Figure 5).
Such a coating may be capable of easily releasing
rigid organisms.

\section*{CONCLUSION}

Synthetic adhesives are efficient because they
involve several mechanisms for dissipating a large
amount of energy upon detachment (see reviews by
Kinloch, 1996, Gay \& Leibler, 1999a, and Creton \&
Fabre, 2002). Some of these mechanisms, and the
associated mechanical aspects, have been reviewed
and may enhance understanding of the settlement
and release of fouling organisms and thereby lead to
improvements in the design of fouling release and
non-adhesive coatings.

\section*{Acknowledgements}

I acknowledge fruitful discussions and collabo-
rations on adhesion phenomena over the past
few years, with Arnaud Chiche, Ioulia Chikina,
Costantino Creton, Guillaume de Crevoisier, Pascale
Fabre, Gwendal Josse, Ludwik Leibler, Fr\'ed\'eric
Nallet, Sylwia Poivet, Elie Rapha\"el, Didier Roux. I
gratefully thank the organizers of the Conference for
the invitation to give a presentation.

\section*{References}

Autumn K, Chang W-P, Fearing R, Hsieh T, Kenny T, Liang L,
Zesch W, Full R J (2000) Adhesive force of a single gecko foot-
hair. {\em Nature (Lond)} {\bf 405}: 681--685

Barthlott W, Neinhuis C (1997) Purity of the sacred lotus, or escape
from contamination in biological surfaces. {\em Planta} {\bf 202}: 1--8

Bico J, Marzolin C, Qu\'er\'e D (1999) Pearl drops. {\em Europhys Lett} {\bf 47}:
220--226

Charmet J-C, Verjus C, Barquins M (1995) Sur la dimension du
contact et la cin\'etique de roulement d’un cylindre long et
rigide sous la surface plane et lisse d’un massif de caoutchouc
souple. {\em C R Acad Sci S\'erie II} {\bf 321}: 443--450

Creton C, Leibler L (1996) How does tack depend on time of
contact and contact pressure? {\em J Polym Sci B} {\bf 34}: 545--554

Creton C, Fabre P (2002) Tack of PSA’s. In: Dillard D A, Pocius A V
(eds) {\em Adhesion Science and Engineering}, Volume 1, {\em The
Mechanics of Adhesion}. Elsevier (In press)

de Crevoisier G, Fabre P, Corpart J-M, Leibler L (1999) Switchable
tackiness and wettability of liquid-crystalline polymers.
{\em Science} {\bf 285}: 1246

Dahlquist C A (1969) Pressure sensitive adhesives. In: Patrick R L
(ed) {\em Treatise on Adhesion and Adhesives}, Volume 2.
Marcel Dekker, New York, pp 219--260

Francis B A, Horn R G (2001) Apparatus-specific analysis of fluid
adhesion measurements. {\em J Appl Phys} {\bf 89}: 4167--4174

Fuller K N G, Tabor D (1975) The effect of surface roughness
on the adhesion of elastic solids. {\em Proc R Soc Lond A} {\bf 345}:
327--342

Gay C, Leibler L (1999a) On stickiness. {\em Phys Today} {\bf 52}: 48--52

Gay C, Leibler L (1999b) Theory of tackiness. {\em Phys Rev Lett} {\bf 82}:
936--939

Gent A N, Schultz J (1972) Effect of wetting liquids on the
strength of adhesion of viscoelatic materials. {\em J Adhesion} {\bf 3}:
281--294

Ghatak A, Chaudhury M K, Shenoy V, Sharma A (2000)
Meniscus instability in a thin elastic film. {\em Phys Rev Lett} {\bf 85}:
4329--4332

Greenwood J A, Williamson J B P (1966) Contact of nominally flat
surfaces. {\em Proc R Soc Lond A} {\bf 295}: 300--319

Griffith A A (1920) The phenomena of rupture and flow in solids.
{\em Philos Trans R Soc Lond A} {\bf 221}: 163--198

Israelachvili J (1992) {\em Intermolecular and Surface Forces}, 2nd Edition.
Academic Press, London \& New York

Kinloch A J (1996) Sticking up for adhesives. {\em Proc R Inst GB} {\bf 67}:
193--217

Lake G J, Thomas A G (1967) The strength of highly elastic
materials. {\em Proc R Soc Lond A} {\bf 300}: 108--115

Lakrout A, Sergot P, Creton C (1999) Direct observation of
cavitation and fibrillation in a probe tack experiment on
model acrylic pressure-sensitive-adhesives. {\em J Adhesion} {\bf 69}:
307--359

Neinhuis C, Barthlott W (1997) Characterization and distribution
of water-repellent, self-cleaning plant surfaces. {\em Ann Bot} {\bf 79}:
667--677

Rapha\"el E, de Gennes P-G (1992) Rubber-rubber adhesion with
connector molecules. {\em J Phys Chem} {\bf 96}: 4002--4007

Saffman P G, Taylor G I (1958) The penetration of a fluid into a
porous medium or Hele-Shaw cell containing a more viscous
liquid. {\em Proc R Soc Lond A} {\bf 245}: 312--329

Schultz M P, Finlay J A, Callow M E, Callow J A (2003) Three
models to relate detachment of low form fouling at laboratory
and ship scale. {\em Biofouling} {\bf 19}(suppl.): S17--S26

Shull K R, Flanigan C M, Crosby A J (2000) Fingering instabili-
ties of confined elastic layers in tension. {\em Phys Rev Lett} {\bf 84}:
3057--3060

Wagner T, Neinhuis C, Barthlott W (1996) Wettability and
contaminability of insect wings as a function of their surface
sculptures. {\em Acta Zool} {\bf 77}: 213--225

Zosel A (1985) Adhesion and tack of polymers: influence of
mechanical properties and surface tensions. {\em Colloid Polymer
Sci} {\bf 263}: 541--553

\end{document}